# The Dirac Algebra and Grand Unification

## Peter Rowlands* and J. P. Cullerne†

*IQ Group and Science Communication Unit, Department of Physics, University of Liverpool, Oliver Lodge Laboratory, Oxford Street, P.O. Box 147, Liverpool, L69 7ZE, UK.   e-mail prowl@hep.ph.liv.ac.uk and prowl@csc.liv.uk

†IQ Group, Department of Computer Science, University of Liverpool, Chadwick Laboratory, Peach Street, Liverpool, L69 72F, UK.

*Abstract*. A representation of the Dirac algebra, derived from first principles, can be related to the combinations of unit charges which determine particle structures. The algebraic structure derives from a broken symmetry between 4-vectors and quaternions which can be applied to the broken symmetry between the three nongravitational interactions. The significance of this relation for Grand Unification is derived by explicit calculation of the running values of the fine structure constants, with suggestions for the calculation of particle masses.

**1 The origin of the Dirac algebra**

Four quantities, and four only, are fundamental in physics: space, time, mass and charge. It seems that, for epistemological reasons, these are structured to form a Klein-4 group in which each has properties which form symmetrical opposites to those of other members of the group, according to the following scheme:[1-6]

| | | | |
|---|---|---|---|
| **space** | real | nonconserved | divisible |
| **time** | imaginary | nonconserved | indivisible |
| **mass** | real | conserved | indivisible |
| **charge** | imaginary | conserved | divisible |

One consequence of this is that we can construct a representation of the four interactions known in nature using a quaternion structure for the sources, with the imaginary units (*i*, *j*, *k*) being assigned to the sources of the strong, electromagnetic and weak interactions (the charge components, $s$, $w$, $e$) and the real or scalar unit (1) being assigned to the source of the gravitational interaction (mass or mass-energy, $m$). By symmetry, we create a 4-vector structure for space and time, with the real or vector units (**i**, **j**, **k**) corresponding to the three dimensions of space and the imaginary or pseudoscalar unit (*i*) corresponding to time. However, for *absolute* symmetry between mass-charge and space-time, the space-time 4-vector must be multivariate or quaternion-like. That is, the vector part must obey the multiplication rule for multivariate vectors (C4 Clifford algebra) in which vectors **a** and **b** have a full product defined by

$$\mathbf{ab} = \mathbf{a.b} + i\, \mathbf{a} \times \mathbf{b} \ .$$

Far from being a problem, this turns out to be physically fruitful, as, using the full product, rather than the scalar product, in quantum mechanical equations leads automatically to the concept of spin without any ad hoc assumptions.[7-8] It also allows us to define simple products for unit vector components which are parallel to those for unit quaternions:

| *vector units* | *quaternion units* |
|---|---|
| $\mathbf{i}^2 = 1$ | $i^2 = -1$ |
| $\mathbf{ij} = -\mathbf{ji} = i\mathbf{k}$ | $ij = -ji = k$ |
| $\mathbf{j}^2 = 1$ | $j^2 = -1$ |
| $\mathbf{jk} = -\mathbf{kj} = i\mathbf{i}$ | $jk = -kj = i$ |
| $\mathbf{k}^2 = 1$ | $k^2 = -1$ |
| $\mathbf{ki} = -\mathbf{ki} = i\mathbf{j}$ | $ki = -ik = j$ . |

The great advantage of the method occurs when we combine the two algebras. Immediately we generate a 32-part algebra, which has the exact properties we require for the algebra of the Dirac equation, and, using this algebra, we can effectively dispense with matrices, which are a much more restricted way of doing quantum mechanics.[9-15] We can express the relationship of this 32-part algebra to that produced by the Dirac $\gamma$ matrices as follows:

$\gamma^0 = i\mathbf{k}, \gamma^1 = i\mathbf{i}, \gamma^2 = i\mathbf{j}, \gamma^3 = i\mathbf{k}, \gamma^5 = i\mathbf{j},$

$\gamma^0\gamma^1 = ij\mathbf{i}, \gamma^0\gamma^2 = ij\mathbf{j}, \gamma^0\gamma^3 = ij\mathbf{k}, \gamma^0\gamma^5 = \mathbf{i}, \gamma^1\gamma^2 = -i\mathbf{k},$
$\gamma^1\gamma^3 = i\mathbf{j}, \gamma^1\gamma^5 = ik\mathbf{i}, \gamma^2\gamma^3 = -i\mathbf{i}, \gamma^2\gamma^5 = ik\mathbf{j}, \gamma^3\gamma^5 = ik\mathbf{k},$

$\gamma^0\gamma^1\gamma^2 = k\mathbf{k}, \gamma^0\gamma^1\gamma^3 = -k\mathbf{j}, \gamma^0\gamma^1\gamma^5 = \mathbf{i}, \gamma^0\gamma^2\gamma^3 = k\mathbf{i}, \gamma^0\gamma^2\gamma^5 = \mathbf{j},$
$\gamma^0\gamma^3\gamma^5 = \mathbf{k}, \gamma^1\gamma^2\gamma^3 = -i\mathbf{i}, \gamma^1\gamma^2\gamma^5 = j\mathbf{k}, \gamma^1\gamma^3\gamma^5 = -j\mathbf{j}, \gamma^2\gamma^3\gamma^5 = j\mathbf{i},$

$\gamma^0\gamma^1\gamma^2\gamma^3 = \mathbf{j}, \gamma^0\gamma^1\gamma^2\gamma^5 = -ii\mathbf{k}, \gamma^0\gamma^1\gamma^3\gamma^5 = ii\mathbf{j}, \gamma^0\gamma^2\gamma^3\gamma^5 = -ii\mathbf{i}, \gamma^1\gamma^2\gamma^3\gamma^5 = \mathbf{k},$

$\gamma^0\gamma^1\gamma^2\gamma^3\gamma^5 = -i.$

From this, we can go on to derive the Dirac equation from the relativistic equation for energy-momentum conservation,

$$E^2 - p^2 - m^2 = 0 \ ,$$

by, first factorizing and attaching the exponential term $e^{-i(Et - \mathbf{p}.\mathbf{r})}$, so that

$$(\pm k E \pm i\mathbf{i}\, \mathbf{p} + i\mathbf{j}\, m)\, (\pm k E \pm i\mathbf{i}\, \mathbf{p} + i\mathbf{j}\, m)\, e^{-i(Et - \mathbf{p}.\mathbf{r})} = 0,$$

and then replacing $E$ and $\mathbf{p}$ in the first bracket with the quantum operators, $i\partial/\partial t$ and $-i\nabla$, to give

$$\left(\pm i k \frac{\partial}{\partial t} \pm i\nabla + i\mathbf{j} m\right)(\pm k E \pm i\mathbf{i}\, \mathbf{p} + i\mathbf{j}\, m)\, e^{-i(Et - \mathbf{p}.\mathbf{r})} = 0.$$

This can be written in the form

$$\left(\pm i k \frac{\partial}{\partial t} \pm i\nabla + i\mathbf{j} m\right)\psi = 0,$$

where the wavefunction

$$\psi = (\pm k E \pm i\mathbf{i}\, \mathbf{p} + i\mathbf{j}\, m)\, e^{-i(Et - \mathbf{p}.\mathbf{r})}.$$

Because of the multivariate nature of $\mathbf{p}$, the second term here may represent either vector $\mathbf{p}$ or scalar $p = \mathbf{1}.\mathbf{p}$, in either a general or a specific defined direction, and consequently, also, $\sigma.\mathbf{p}$. (The same options are available to $\nabla$.) The four sign options for the $E$ and $\mathbf{p}$ terms in the anticommuting pentad $(\pm k E \pm i\mathbf{i}\, \mathbf{p} + i\mathbf{j}\, m)$, are mostly conveniently incorporated by representing $\psi$ as a 4-term column vector, with the differential operator as the equivalent 4-term row vector. The exponential term is common to all four solutions. It can be easily shown that these solutions are equivalent to those incorporated in the conventional Dirac spinor, and the conventional results of Dirac theory, such as the energy states of the hydrogen atom, are easily generated in a compact algebraic form. In addition to this, we can derive explicit wavefunctions for scalar and vector bosons, Bose-Einstein condensates, free fermions and baryons, with their respective parity states; and annihilation, creation and vacuum operators. Other topics easily dealt with using this formalism include C, P and T transformations; Pauli exclusion; propagators; and quantum field integrals.

This is all very convenient, but does it have a more fundamental significance? The important fact here is that we have generated our Dirac algebra from a mathematical representation of charge. Is it related physically? The answer is that it most certainly is. First, let us look again at the algebra. The five composite terms of the anticommuting pentad ($i\mathbf{k}$, $i\mathbf{i}$, $i\mathbf{j}$, $i\mathbf{k}$, $j$, in the most fundamental representation) turn out to be more fundamental in generating the whole algebra than the eight primary terms that we began with ($i$, $\mathbf{i}$, $\mathbf{j}$, $\mathbf{k}$, 1, $i$, $j$, $k$). Mathematically, also, the process of creating such a pentad is highly restricted. It involves removing one of the 3-dimensional subsets (charge or space) and imposing it upon the rest, and it is this process which creates the Dirac state.

If we take the 3-dimensional parameter to be removed as charge, we effectively end its existence as a separate entity by combining its three components with those, respectively of time, space and mass, the algebraic rules requiring the preservation of the other 3-dimensional unit (space) as a complete entity. In doing so, we *create* the three composite quantities that form the Dirac state – energy, momentum and rest mass ($E$, **p**, $m$) – and these, like the three components of charge, which they effectively incorporate, become separately quantized and conserved. And, though the separate charges are no longer explicit, we will see that they are still present in a hidden form through the unifying concept of angular momentum.

Why is this relevant to Grand Unification? The answer is that it is in the *separate* ways that the charges represent angular momentum that the broken symmetry between the three nongravitational forces manifests itself. We will see that, in this broken symmetry, $w$ aligns itself with the $E$ term, $s$ with the **p** term, and $e$ with the $m$ term. And it is in the combination of these as an anticommuting pentad that we get the first idea of how an $SU(5)$ type of Grand Unification can come about. An extension to $U(5)$ suggests a way in which gravity, also, can be incorporated, as our calculations for the unification of the other three forces would seem to imply.

**2 The strong interaction: $SU(3)$**

Our aim here is to show that the $SU(3)$ structure for the strong interaction is a direct result of its existence in the Dirac algebra. We will then perform the same derivation for the electroweak $SU(2)_L \times U(1)$. The $SU(3)$ property comes from the three-dimensionality of **p**. In the Dirac state vector, an angular momentum state must remain unspecified as to direction, although one direction (and one only) may be well defined. This gives us two ways of constructing a fermion wavefunction. We can either specify the three components of the angular momentum, and allow the coexistence of three directional states as long as none is specified, or we can specify the *total* angular momentum as well defined in a single direction (though without specific preference). In the first case, we have a 'baryon' structure with a quaternion state vector of the form:

$$(k E \pm i\boldsymbol{i}\, p_1 + i\boldsymbol{j}\, m)\ (k E \pm i\boldsymbol{i}\, p_2 + i\boldsymbol{j}\, m)\ (k E \pm i\boldsymbol{i}\, p_3 + i\boldsymbol{j}\, m)\ ,$$

with six degrees of freedom for the spin ($\pm p_1$, $\pm p_2$, $\pm p_3$), which we can equate with six coexisting representations for the 'three colour' or 'three quark' combinations. In the second case, we have a 'free fermion' (or lepton) structure with a quaternion state vector of the form:

$$(k E \pm i\boldsymbol{i}\, \mathbf{p} + i\boldsymbol{j}\, m)\ .$$

Clearly, these have the same overall structure when we equate **p** successively with $\pm p_1$, $\pm p_2$, $\pm p_3$. The baryon structure is determined solely by the nilpotent nature of the fermion wavefunction (the fact that it is a square root of zero). Putting in an extra **p** into the brackets missing them, would immediately reduce the state vector to zero.

Now, the standard QCD representation of the baryon is the antisymmetric colour singlet of $SU(3)$:
$$\psi \sim (BGR - BRG + GRB - GBR + RBG - RGB),$$

which gives us a mapping of the form:

$$
\begin{array}{ll}
BGR & (k E + i j\, m)\, (k E + i j\, m)\, (k E + i i\, \mathbf{p} + i j\, m) \\
-BRG & (k E + i j\, m)\, (k E - i i\, \mathbf{p} + i j\, m)\, (k E + i j\, m) \\
GRB & (k E + i j\, m)\, (k E + i i\, \mathbf{p} + i j\, m)\, (k E + i j\, m) \\
-GBR & (k E + i j\, m)\, (k E + i j\, m)\, (k E - i i\, \mathbf{p} + i j\, m) \\
RBG & (k E + i i\, \mathbf{p} + i j\, m)\, (k E + i j\, m)\, (k E + i j\, m) \\
-RGB & (k E - i i\, \mathbf{p} + i j\, m)\, (k E + i j\, m)\, (k E + i j\, m)\,, \quad (1)
\end{array}
$$

with each term equivalent to $-p^2(kE + ii\,\mathbf{p} + ij\,m)$ or $-p^2(kE - ii\,\mathbf{p} + ij\,m)$, representing three cyclic and three anticyclic combinations. Because there is only one spin term, this representation predicts that the spin is a property of the baryon wavefunction as a whole, not of component quark wavefunctions.

With the spinor terms included, each of these is represented by a tensor product of three spinors, for example:

$$(kE + ij\, m)\,(kE + ij\, m)\,(kE + ii\,\mathbf{p} + ij\, m)\left(\tfrac{1}{2}\right)\otimes\left(\tfrac{1}{2}\right)\otimes\left(\tfrac{1}{2}\right)$$

where

$$\left(\tfrac{1}{2}\right)\otimes\left(\tfrac{1}{2}\right)\otimes\left(\tfrac{1}{2}\right) = \left(\tfrac{3}{2}\right)\oplus\left(\tfrac{1}{2}\right)\oplus\left(\tfrac{1}{2}\right)$$

So the representation encompasses both spin ½ and spin 3/2 baryon states.

In conventional $SU(3)$ theory, the baryon structure is maintained by a strong interaction between the three component (quark) states, maintained by an exchange of massless gluons. The $SU(3)$ symmetry for this strong source is conventionally expressed using a covariant derivative of the form:

$$\partial_\mu \to \partial_\mu + ig_s \frac{\lambda^\alpha}{2} A^{\alpha\mu}(x)\,.$$

In component form:

$$ip_1 = \partial_1 \to \partial_1 + ig_s \frac{\lambda^\alpha}{2} A^{\alpha 1}(x)$$

$$ip_2 = \partial_2 \to \partial_2 + ig_s \frac{\lambda^\alpha}{2} A^{\alpha 2}(x)$$

$$ip_3 = \partial_3 \to \partial_3 + ig_s \frac{\lambda^\alpha}{2} A^{\alpha 3}(x)$$

$$E = i\partial_0 \to i\partial_0 - g_s \frac{\lambda^\alpha}{2} A^{\alpha 0}(x)\,.$$

Using this, we may observe that the baryon state vector has the same form as the eigenvalue of the Dirac differential operator, which is the product of the three terms:

$$\left( k\left( E - g_s \frac{\lambda^\alpha}{2} A^{\alpha 0} \right) \pm i\left( \partial_1 + ig_s \frac{\lambda^\alpha}{2} A^{\alpha 1} \right) + ij\, m \right)$$

$$\left( k\left( E - g_s \frac{\lambda^\alpha}{2} A^{\alpha 0} \right) \pm i\left( \partial_2 + ig_s \frac{\lambda^\alpha}{2} A^{\alpha 2} \right) + ij\, m \right)$$

$$\left( k\left( E - g_s \frac{\lambda^\alpha}{2} A^{\alpha 0} \right) \pm i\left( \partial_3 + ig_s \frac{\lambda^\alpha}{2} A^{\alpha 3} \right) + ij\, m \right).$$

The nilpotent nature of the term ($kE \pm i i\, \mathbf{p} + ij\, m$) ensures that the only way of preserving nonzero fermionic structure here is to write this expression in one of the forms:

$$\left( k\left( E - g_s \frac{\lambda^\alpha}{2} A^{\alpha 0} \right) \pm i\left( \partial_1 + ig_s \frac{\lambda^\alpha}{2} \mathbf{A}^\alpha \right) + ij\, m \right)\left( k\left( E - g_s \frac{\lambda^\alpha}{2} A^{\alpha 0} \right) + ij\, m \right)\left( k\left( E - g_s \frac{\lambda^\alpha}{2} A^{\alpha 0} \right) + ij\, m \right)$$

$$\left( k\left( E - g_s \frac{\lambda^\alpha}{2} A^{\alpha 0} \right) + ij\, m \right)\left( k\left( E - g_s \frac{\lambda^\alpha}{2} A^{\alpha 0} \right) \pm i\left( \partial_1 + ig_s \frac{\lambda^\alpha}{2} \mathbf{A}^\alpha \right) + ij\, m \right)\left( k\left( E - g_s \frac{\lambda^\alpha}{2} A^{\alpha 0} \right) + ij\, m \right)$$

$$\left( k\left( E - g_s \frac{\lambda^\alpha}{2} A^{\alpha 0} \right) + ij\, m \right)\left( k\left( E - g_s \frac{\lambda^\alpha}{2} A^{\alpha 0} \right) + ij\, m \right)\left( k\left( E - g_s \frac{\lambda^\alpha}{2} A^{\alpha 0} \right) \pm i\left( \partial_1 + ig_s \frac{\lambda^\alpha}{2} \mathbf{A}^\alpha \right) + ij\, m \right)$$

which are, of course, parallel to the six forms expressed in (1).

Physically, this means that the carrier of the 'colour' component of the strong force ($ig_s \lambda^\alpha \mathbf{A}^\alpha / 2$) is 'transferred' between the quarks at the same time as the spin, both being incorporated into the **p** term, and the current that effects the 'transfer' is carried by the gluons or generators of the strong field. To ensure that the baryon wavefunction remains noncollapsable, and that the strong interaction remains gauge invariant, all the representations or 'phases' are present at the same time, and equally probable. The three quark 'colours' are no more capable of separation from each other than are the three dimensions of space (and, of course, these conditions are exactly equivalent to each other, as they stem from exactly the same origin). So when we describe the strong force as effecting a 'transfer' of strong charge or 'colour' field, we a really expressing, in a relatively simple way, the fact that the innate gauge invariance of the strong interaction is the same thing as the conservation of the directional aspect of angular momentum.

## 3 The electroweak interaction: $SU(2)_L \times U(1)$

Here, we will first outline the observed patterns of these interactions, and then show how they can be related to the creation of the Dirac state. Experimentally, we find that weak interactions all follow the same pattern. In the case of leptons, it is

$$e + \nu \rightarrow e + \nu. \tag{2}$$

For quarks, it is

$$u + d \rightarrow u + d,$$

with $d$ taking the place of $e$, and $u$ that of $\nu$. For weak interactions involving both leptons and quarks (for example, $\beta$ decay), the pattern is once again the same:

$$d + \nu \rightarrow e + u.$$

Let us, for the moment, consider (2). There are four possible vertices (assuming left-handed components only).

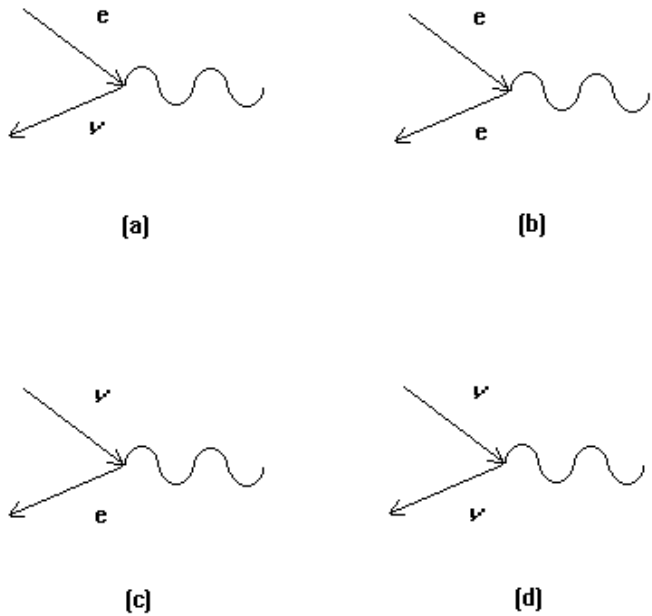

All the vertices must be true at once. So the interaction can be described as a mixing or superposition of the four possibilities. However, the second vertex (b), and this one alone, also represents a possible electromagnetic interaction, giving us a 1 to 4 ratio for the occurrence of the electromagnetic to weak interaction at the energy which the vertices characteristically represent (that of the *W / Z* bosons). This suggests that particle charge structures at this energy must be such that the electroweak mixing ratio,

$$\sin^2\theta_W = \frac{e^2}{w^2} = \frac{\Sigma\, t_3^2}{\Sigma\, Q^2} = 0.25\ .$$

If we take the quaternion state vectors for the fermionic components of the four vertices, we obtain, for the case where the spins of the interacting fermions are assumed parallel (total 0 for fermion-antifermion combination):

(a)          $(k E - i i \mathbf{p} + i j m) \ldots (- k E + i i \mathbf{p}) \ldots = 4m^2$ ;
(b)          $(k E - i i \mathbf{p} + i j m) \ldots (- k E + i i \mathbf{p} + i j m) \ldots = 4m^2$ ;
(c)          $(k E - i i \mathbf{p}) \ldots (- k E + i i \mathbf{p} + i j m) \ldots = 4m^2$ ;
(d)          $(k E - i i \mathbf{p}) \ldots (- k E + i i \mathbf{p}) \ldots = 4m^2$ .

where $(k E - i i \mathbf{p} + i j m) \ldots$ represents a column or row vector with the terms:

$(k E - i i \mathbf{p} + i j m);\ (k E + i i \mathbf{p} + i j m);\ (- k E + i i \mathbf{p} + i j m);\ (- k E - i i \mathbf{p} + i j m)$ ,

and so on. Applying a standard normalisation, these sums become $m^2 / E^2$, implying that, without an *m* term, all four vertices would become 0. The *m* term must arise from the fact that **p** is not purely composed of left-handed helicity states (with – **p** right-handed), but incorporates a right-handed component, which itself cannot contribute to the weak interaction because of charge-conjugation violation and the presence of a weak filled vacuum. The right-handed component *can only arise from the presence of the electromagnetic interaction*. The weak interaction cannot exist as a pure left-handed interaction, without a mixing with the electromagnetic interaction to produce the necessary non-zero mass through the introduction of right-handed states.

Suppose we now put into the *E* and **p** terms of the state vector the covariant derivatives for the electroweak interaction. The scalar part goes with *E* and the vector part with **p**. Mass is produced by the mixing of *E* with **p** via the relativistic connection between these terms. It is also produced by the mixing of $B^0$ with $W^+$, $W^0$, and $W^-$, which we may now identify with the four vertices (d), (a), (b), and (c). By choosing the single, well-defined direction of spin or angular momentum (**p**) to be, in principle, the one where the total value for the interacting fermion-antifermion combination is 0, we can ensure that the mixing is specifically between the neutral components, $B^0$ and $W^0$, and create one massless *combination* to represent the carrier of the pure electromagnetic interaction ($\gamma$), with the other being the massive neutral weak carrier $Z^0$. The mixing must be such as to define the ratio of the two interactions, $\sin^2\theta_W$, at 0.25. (The other two vertices, $W^+$ and $W^-$, then fulfil the requirements for the existence of states corresponding to total spin values of +1 and – 1.)

Because *m* is determined from the combination of *E* and **p**, we can, by appropriate choice of the value of *m*, make these compatible if we additionally define a combination of the coupling constants related to the $SU(2)_L$ and $U(1)$ symmetries, *g'* and *g*, which removes $B^3$ from *E* and $W^0$ from **p**. Because the combinations of *g'* and *g*, now represent the pure electromagnetic and weak coupling constants, *e* and *w* (= *g*), we must necessarily obtain the ratio $e^2 / w^2 = 0.25$, and both quarks and leptons must be structured to observe this.

## 4 The Dirac interpretation

How does this relate to the Dirac nilpotent structure? In principle, it seems likely that the conservation properties of the weak and electromagnetic charges are determined by those of the angular momentum operator, like those of the strong charge. However, neither of these is attached directly to **p**; one is attached to *E* and one to *m*, and it is the *combination* of these which affects **p**. It is for this reason that we think of the electric and weak forces as being in some way combined. In principle, the charge represented by the quaternion label ***k*** (which we call the weak charge) produces two sign options for *iE*, because the algebra demands complexification of *E*, and there are necessarily two mathematical solutions. Only the positive solution, however, should be physically meaningful for energy, and we compensate by creating a filled vacuum for the ground state of the universe, in which states with negative *E* (or antifermions) would not exist, though they are allowed by the parallel mathematical status of the quaternion labels as square roots of –1, which permits charge conjugation or reversal of the signs of the quaternion labels.

The filled vacuum is specifically a ***k*** or weak vacuum, and it results in a violation of charge conjugation symmetry for the weak interaction, with consequent violation of either time reversal symmetry or parity. Charge conjugation violation effectively means that the weak interaction cannot tell the difference between + and – signs of weak charge, though it can tell the difference between fermions and antifermions. A consequence is that *w* and *s* charges takes only one effective sign for fermions, though charge conjugation should allow two signs for *w* (or *E*) if that of *s* (or **p**) is fixed.

By suppressing the alternative sign for *w* we ensure that quark and free fermion states become mixed states, containing +*w*, and suppressed –*w* states involving respective violations of parity and time reversal symmetry. We find also that only one state of σ.**p** exists for the pure *w* interaction for fermions; because σ = –**1**, this is the state of negative helicity or left-handedness. To create the states of positive helicity or right-handedness, which should also exist for –**p**, we have to introduce mass, which is associated in the Dirac state with the ***j*** quaternion label, which defines what we call the electric charge. The introduction of *m* also introduces the *E* / **p** mixing, which produces a right-handed component mixed with the left-handed. Such a mixing can only be produced by a mixing of the effects of *e* charges with those of *w*.

The presence or absence of *e* charges creates the characteristic $SU(2)_L$ 'isospin' pattern associated with the weak interaction, for this interaction must be both uniquely left-handed for fermion states and indifferent to the presence or absence of the electric charge, which introduces the right-handed element. The $SU(2)$ produces a quantum number, $t_3$, such that $(t_3)^2 = (½)^2$ in half the total number of possible states. For free fermions, with 0 or ±1 as the quantum number for the electric force, and so with $Q^2 = 1$ in half the total number of possible states, the key electroweak mixing parameter becomes $\sin^2\theta_W = \Sigma (t_3)^2 / \Sigma Q^2 = 0.25$, which is the same proportion as would be obtained by taking the electron and neutrino as the possible free fermion states. Since the weak force must also be indifferent to the presence of the strong interaction, or to the directional state of the angular momentum operator, then the same mixing proportion must exist also for quark states, and separately for each colour, so none is preferred.

It should be noted here that *exchange* of electromagnetic charge, through, say, $W^+$ or $W^-$, is nothing to do with the electromagnetic interaction, but is rather an indication that the weak interaction is unable to detect the presence of the electromagnetic charge, that is, that a 'weak interaction' is a statement that all states of a particle with the same weak charge are equally probable, given the appropriate energy conditions, and that gauge invariance is maintained with respect to them. In principle, weak bosons are massive because they act as carriers of the electromagnetic charge, whereas electromagnetic bosons (or photons) are massless because they do not – the quantitative value of the mass must be determined from the coupling of the weak charge to the asymmetric vacuum state which produces the violation of charge conjugation in the weak interaction. The weak interaction is also indifferent to the presence of the strong charge, and so cannot distinguish between quarks and leptons – hence, the intrinsic identity of purely lepton weak interactions with quark-lepton or quark-quark ones – and, in the case of quarks, it cannot tell the difference between a filled 'electromagnetic vacuum' (up quark) and an empty one (down quark). The weak interaction, in addition, is also indifferent to the sign of the weak charge, and responds (via the vacuum) only to the status of fermion or antifermion – hence, the CKM mixing.

**5 The charge structures of quarks and leptons**

The creation of the Dirac state determines the nature of the symmetries applicable to the three interactions. It also determines the distribution of the units of the charges in both quarks and leptons. For example, it is evident that the 'transfer' of the momentum or angular momentum (**p** or σ.**p**) term between the three components of the baryon wavefunction is identical, in principle, to the 'transfer' of a unit of strong charge through the term $ig_s \lambda^\alpha \mathbf{A}^\alpha / 2$. (The fact that the 'transfer' is completely gauge invariant is equivalent to making it occur at a constant rate, equivalent to the action of a constant force, or a potential which increases linearly with distance, as observed with the strong interaction; spherical symmetry and the nilpotent structure then require an additional Coulomb component.[15]) The *e* and *w* units on both quarks and leptons must also be structured in such a way that $\sin^2\theta_W = \Sigma (t_3)^2 / \Sigma Q^2 = 0.25$.

One way of representing the strong charge 'transfer' would be to take the σ.**p** as equivalent in unit charge terms to an expression of the form **1.r**, where **r** is successively the unit vector components **i**, **j**, **k**, and apply this to the strong charge quaternion operator *i*. The units of strong charge then become 0*i* or 1*i*, depending on what we may imagine to be the instantaneous direction of the angular momentum vector, which carries the charge unit, and only one component of the baryon will have this unit at any instant. Of course, because of gauge invariance, there is no such thing as an instantaneous direction, and all possible states will exist at once. A consequence of this is that spin is not intrinsic to the quarks but is a property of the system. A spin direction is uniquely definable only for the baryon as a whole, and not for the component quarks.

Now, it is this same angular momentum term (**p** or σ.**p**), which carries the information concerning the conservation of the other two charge terms, and, in effect, the three charges are separately conserved because they represent three aspects of the angular momentum conservation process. So, we can generalise this procedure to apply to all three charges simultaneously. Here we apply the unit vector components **r**$_1$, **r**$_2$, **r**$_3$, randomly taking the values **i**, **j**, or **k**, to the weak, strong and electromagnetic charges specified by the quaternion labels *k*, *i*, *j*. Taking the scalar product of these terms with a unit vector (**1**), which is equivalent to taking the product σ.**p** (or σ.**p̂**) in the Dirac state, then determines which charges take unit and which zero values.

In the case of the weak and electric charges, the random unit vector components are associated respectively with the sign, and the magnitude of the angular momentum state, through the connections of **p** with *E* and **p** with *m*. In these cases, we are not concerned with the directional components, which is entirely associated with the strong charge. Thus we are able to associate a fixed single vector (**i**, **j**, or **k**, though the choice is arbitrary) with each of the quaternion labels (*k* and *j*) specifying *w* and *e*. In the baryon system, when strong charges are present, we must assume that the vectors assigned to the weak and electric charges are not aligned. This is because, for the weak and electric forces to carry no directional information, the charges and their associated vectors must be arranged for only one of the three quarks in a baryon to be differentiated at any instant, and the *e* and *w* values so specified must be separated.

In the case of the free fermion or lepton, however, the quaternion state vector must necessarily exclude the strong charge or intrinsically directional components of angular momentum. The angular momentum must have a single well-defined direction, and so the random vectors associated with the electric and weak charges must be aligned (to specify the direction). In fact, alignment of these vectors can be taken as the signature of a free fermion, excluding the strong interaction. In the two possible weak isospin states, we then have the electric charge unit as 0 or –*e* (the negative value being adopted by convention), with the weak unit as effectively *w* (either directly or by violation of parity or time reversal symmetry). From its application purely to left-handed states, we obtain $\sin^2\theta_W = \Sigma (t_3)^2 / \Sigma Q^2 = 0.25$.

Applying this now to quarks, we have the same weak isospin states for one 'colour', but we also find that the only corresponding isospin states for the other colours that retain both the accepted value of $\sin^2\theta_W$ and the variation of only one quark 'instantaneously' in three, are *e* and 0. In effect, this is like adding a full *e e e* background or 'vacuum' to the original 0 0 –*e*, so that the two states of weak isospin in the three colours become:

$$\begin{array}{ccc} e & e & 0 \\ 0 & 0 & -e \end{array}.$$

The creation of three generations, as well as isospin states, results from the violations of parity and time reversal symmetry which are consequent upon the effective suppression of –*w* states for fermions. We can represent charge conjugation violation in one of two forms by $z_P$ and $z_T$, depending on whether it comes with *P* or *T* violation. These are not algebraic operators, but symbols which say that, in treating the *w* of the second and third generator as though it were positive in the same way as the *w* of the first generation, we have to violate charge conjugation symmetry. We can now express the result algebraically as follows:

| | |
|---|---|
| down | $-j\mathbf{r}_1 + i\mathbf{r}_2 + k\mathbf{r}_3$ |
| up | $-j(\mathbf{r}_1 - 1) + i\mathbf{r}_2 + k\mathbf{r}_3$ |
| strange | $-j\mathbf{r}_1 + i\mathbf{r}_2 + z_P k\mathbf{r}_3$ |
| charmed | $-j(\mathbf{r}_1 - 1) + i\mathbf{r}_2 + z_P k\mathbf{r}_3$ |
| bottom | $-j\mathbf{r}_1 + i\mathbf{r}_2 + z_T k\mathbf{r}_3$ |
| top | $-j(\mathbf{r}_1 - 1) + i\mathbf{r}_2 + z_T k\mathbf{r}_3$ |

Here, $-j$ represents electric charge (traditionally negative), *i* strong, *k* weak. Each term is successively scalar multiplied by the unit vectors **i**, **j**, and **k** (totalling **1**) to produce the component 'coloured' quarks of the composite baryons; each of **i**, **j**, **k** representing one colour of quark, **1** the composite particle. The antiparticles simply reverse all the signs. For the corresponding leptons (where $\mathbf{r}_1 = \mathbf{r}_3$ and there is no $\mathbf{r}_2$ term), we have:

| | |
|---|---|
| electron | $-j\mathbf{r}_1 + k\mathbf{r}_1$ |
| *e* neutrino | $-j(\mathbf{r}_1 - 1) + k\mathbf{r}_1$ |
| muon | $-j\mathbf{r}_1 + z_P k\mathbf{r}_1$ |
| $\mu$ neutrino | $-j(\mathbf{r}_1 - 1) + z_P k\mathbf{r}_1$ |
| tau | $-j\mathbf{r}_1 + z_T k\mathbf{r}_1$ |
| $\tau$ neutrino | $-j(\mathbf{r}_1 - 1) + z_T k\mathbf{r}_1$ |

It is significant that, even though $\mathbf{r}_1$, $\mathbf{r}_2$, and $\mathbf{r}_3$ are completely random when treated separately, the number of different outcomes is reduced by repetitions, and is five, as in the anticommuting pentads of the Dirac algebra, rather than, say, 27; and effectively we 'privilege' one of $\mathbf{r}_1$, $\mathbf{r}_2$, $\mathbf{r}_3$ by allowing it complete variation with respect to the others ($\mathbf{r}_2$ being the one selected). This is effectively the same as 'privileging' **p** (or **p̂**) as a vector term with full variation in the Dirac anticommuting pentad. 27 degrees of freedom are thus reduced to 5 because, though $\mathbf{r}_1$, $\mathbf{r}_2$ and $\mathbf{r}_3$ are independent, in principle, there are only 5 patterns of unit and zero charges resulting.

The charge conjugation represented by $z_P$ or $z_T$ is brought about by the filled weak vacuum needed to avoid negative energy states; the terms ($\mathbf{r}_1 - \mathbf{1}$) and $\mathbf{r}_1$, which represent the two states of weak isospin (the –1, of course, really represents +1 if *j* is conventionally negative), are associated with this idea. In a sense the **1** is a 'filled' state, while 0 is an unfilled state. We are, thus, creating two possible vacuum states to allow variation of the sign of electric charge by weak isospin, and linking this variation to the filling of the vacuum which occurs in the weak interaction. The weak and electric interactions are linked by this filled vacuum in the $SU(2)_L \times U(1)$ model, as they are here by our description of weak isospin, and the $SU(2)_L$ comes from the two states of weak isospin in the charge-conjugation violated (hence left-handed) case.

Symmetry-breaking is, in fact, a clear consequence, of the setting up of this algebraic model. When time, space and mass map onto the charges *w*-*s*-*e*, to create the anticommuting Dirac pentad, only one of the charges (*s*) has the full range of vector options. If we 'fix' one of the others (say *e*) for *s* to vary against, then there are only 2 remaining options for *w*, unit on the same colour as *e* or unit on a different one. We can refer to this as *w* 'on' and 'off' *e*. In fact, if the full variation of *s* is to be allowed, and the combination of *w*, *s*, *e* all 'on' is forbidden, because it denies the necessary three degrees of freedom in the direction of angular momentum, then the combination of *w* and *e* both 'on' can only happen in the absence of *s*, which is what we describe as the lepton state. The reason why we fix *e*, of course, rather than *w*, is because the mapping has made *e* mass-like, and *w* time-like. The time-like *w* has two mathematical states (like *T* or *E*), while *e* has one (like *m*).

The weak interaction can be thought of as a swapping of *w* from *e* 'on' to *e* 'off' or vice versa, creating the $SU(2)_L$, but, in fact, there is no mechanism for doing this directly, as there is for rotation in the strong interaction, because there is no combined system of coexisting states to make it possible. What we *can* do, however, is to annihilate and create, and instead of swapping over *w*, we annihilate and create *e*, either filling the vacuum or emptying it. However, we cannot annihilate or create a charge without also annihilating or creating its antistate, and the weak interaction (unlike the strong) always involves the equivalent of particle + antiparticle = particle + antiparticle, or a double particle interaction going both ways at once. We don't know which it *really* is because the weak interaction works to *prevent* such knowledge. It is because of the filling and emptying of the vacuum via the *e* charge that rest mass is involved. $W^+$ and $W^-$ involve a one-way *e* transition, $Z^0$ involves a two-way *e* transition, the purely electromagnetic (*U*(1)) $\gamma$ no *e* transition. This gives us 0.25 for the electric / weak ratio. The same value also occurs for the weak isospin quantum number squared.

**6 Han-Nambu quark model**

It may be evident that our algebraic scheme for particle charge structures is essentially an extension of the Han-Nambu integrally-charged coloured quark model,[16] which was largely discarded for general use (though not entirely in principle) before the parallel phenomenon of the fractional quantum Hall effect was discovered and explained in condensed matter physics.[17] Laughlin, who explained the fractional quantum Hall effect as resulting from a single fermion forming a bosonic-type state with an odd number of magnetic flux lines, has recently hinted at its relevance for explaining fractional charges in particle physics. In his 1998 Nobel Lecture, he states the following opinion: 'The fractional quantum Hall effect is fascinating for a long list of reasons, but it is important in my view primarily for one: It establishes experimentally that both particles carrying an exact fraction of the electron charge *e* and powerful gauge forces between these particles, two central postulates of the standard model of elementary particles, can arise spontaneously as emergent phenomena. Other important aspects of the standard model, such as free fermions, relativity, renormalizability, spontaneous symmetry breaking, and the Higgs mechanism, already have apt solid-state analogues and in some cases were even modeled after them, but fractional quantum numbers and gauge fields were thought to be fundamental, meaning that one had to postulate them. This is evidently not true.'[18]

There seems, in fact, to be no reason why, in a *fully* gauge invariant theory of the strong interaction, in which the quark colours are intrinsically inseparable, the underlying charges could not be integral while always being perceived as fractional in effect, exactly as in the Laughlin version of the fractional quantum Hall effect. The integral model for fundamental charges thus become just another way of representing the theory of experimentally-observed fractional charges. As Frank Close expressed it in 1979: 'Imagine what would happen if the colour nonsinglets were pushed up to infinite masses. Clearly only colour **1** [singlets] would exist as physically observable states and quarks would in consequence be permanently confined. At any finite energy we would only see the "average" quark changes and phenomenonologically we could not distinguish this from the Gell-Mann model where the quarks form three identical triplets.'[19] The fractional charges would not, in fact, even be 'averages'; they would be exact because of the effectively infinite rate of 'rotation' between the coloured states or phases. With *perfect* gauge invariance for the strong interaction, as we have always assumed, and hence perfect infrared slavery, then there will be no transition regime between implicit and explicit colour; there will be no finite energy range at which integral charges or colour properties will emerge. The charges will be exactly distributed between the quark components of baryons and mesons, and will be exactly fractional, in every way identical to the fractional charges in the standard theory.

Our algebraic model suggests that we can represent the behaviour of fundamental charges in terms of a set of 'quark' tables, A-E, [1,3,20-24] which are shown, in reduced form, below:

**A**

|   |     | **B** | **G** | **R** |
|---|-----|-------|-------|-------|
| u | +*e* | 1*j* | 1*j* | 0*i* |
|   | +*s* | 1*i* | 0*k* | 0*j* |
|   | +*w* | 1*k* | 0*i* | 0*k* |
|   |     |       |       |       |
| d | −*e* | 0*j* | 0*k* | 1*j* |
|   | +*s* | 1*i* | 0*i* | 0*k* |
|   | +*w* | 1*k* | 0*j* | 0*i* |
|   |     |       |       |       |

**B**

|   |     | **B** | **G** | **R** |
|---|-----|-------|-------|-------|
| u | +*e* | 1*j* | 1*j* | 0*k* |
|   | +*s* | 0*i* | 0*k* | 1*i* |
|   | +*w* | 1*k* | 0*i* | 0*j* |
|   |     |       |       |       |
| d | −*e* | 0*i* | 0*k* | 1*j* |
|   | +*s* | 0*j* | 0*i* | 1*i* |
|   | +*w* | 1*k* | 0*j* | 0*k* |
|   |     |       |       |       |

**C**

|   |     | **B** | **G** | **R** |
|---|-----|-------|-------|-------|
| u | +*e* | 1*j* | 1*j* | 0*k* |
|   | +*s* | 0*i* | 1*i* | 0*j* |
|   | +*w* | 1*k* | 0*k* | 0*i* |
|   |     |       |       |       |
| d | −*e* | 0*j* | 0*k* | 1*j* |
|   | +*s* | 0*i* | 1*i* | 0*k* |
|   | +*w* | 1*k* | 0*j* | 0*i* |
|   |     |       |       |       |

**D**

|   |     | **B** | **G** | **R** |
|---|-----|-------|-------|-------|
| u | +*e* | 1*j* | 1*j* | 0*i* |
|   | +*s* | 0*k* | 1*i* | 0*j* |
|   | +*w* | 0*i* | 0*k* | 1*k* |
|   |     |       |       |       |
| d | −*e* | 0*i* | 0*k* | 1*j* |
|   | +*s* | 0*j* | 1*i* | 0*i* |
|   | +*w* | 0*k* | 0*j* | 1*k* |
|   |     |       |       |       |

**E**

|   |     | **B** | **G** | **R** |
|---|-----|-------|-------|-------|
| u | +*e* | 1*j* | 1*j* | 0*j* |
|   | +*s* | 0*k* | 0*i* | 1*i* |
|   | +*w* | 0*i* | 0*k* | 1*k* |
|   |     |       |       |       |
| d | −*e* | 0*i* | 0*k* | 1*j* |
|   | +*s* | 0*j* | 0*i* | 1*i* |
|   | +*w* | 0*k* | 0*j* | 1*k* |
|   |     |       |       |       |

We can use the fact that the charges are irrotational, but the quaternions are not, to derive the essential features of the Standard Model. Even in this case, E appears to be excluded by requiring all three quaternions to be attached to specified charges (losing the three required degrees of freedom, and, at the same time, necessarily violating Pauli exclusion). If applied to known fermions, it would appear that A-C must represent the coloured quark system, with *s* pictured as being 'exchanged' between the three states (although, of course, in reality, all the states exist at once), while D-E, with the exclusion of the *s* charge, represent leptons. The antifermions are generated by reversing all charge states, while two further generations are required by the exclusion of negative values of *w* in fermion states by the respective violations of parity and time reversal symmetry.

**7 *SU*(5) / *U*(5) Grand Unification**

Our demonstration that the 5-fold Dirac algebra is responsible for the symmetry breaking which leads to the $SU(3) \times SU(2)_L \times U(1)$ splitting in the interactions between fundamental particles, suggests that Grand Unification may indeed involve the *SU*(5) group, as is currently believed, though not in the form of the minimal *SU*(5) theory, originally proposed.[25] In principle, we derive five representations of the electric, strong and weak charge states (A-E), which map onto the charge units (*e*, *s*, *w*), and the five quantities (*m*, **p**, *E*) involved in the Dirac equation. The 24 *SU*(5) generators can be represented in terms of any of these units, for example, in the form:

|       | $\bar{s}_G$ | $\bar{s}_B$ | $\bar{s}_R$ | $\bar{w}$ | $\bar{e}$ |
|-------|-------------|-------------|-------------|-----------|-----------|
| $s_G$ |             |             |             |           |           |
| $s_B$ |             | Gluons      |             | Y         | X         |
| $s_R$ |             |             |             |           |           |
| $w$   |             | Y           |             | $Z^0, \gamma$ | $W^-$ |
| $e$   |             | X           |             | $W^+$     | $Z^0, \gamma$ |

The minimal *SU*(5) model, of course, disregards what would be the 25[th] generator in a *U*(5) representation on the grounds that it is not observed. However, if such a particle existed, it would couple to all matter in proportion to the amount, and, as a colour singlet, would be ubiquitous. This is precisely what we need for gravity, and, if we can show that gravity is significant to the Grand Unification of the electromagnetic, strong and weak forces, then we might well be entitled to put forward the *U*(5) group as the true Grand Unification group. This would have the advantage of making all the generators become pure phases, and identical in form, at the Grand Unification energy.

# 8 Grand Unification calculations and predictions

At present, the current 'best fit' theory (minimal $SU(5)$) does not make satisfactory predictions for Grand Unification. It predicts a Grand Unification energy of order $10^{15}$ GeV, about four orders below the Planck mass ($M_P$), at which quantum gravity becomes significant. However, this is only a compromise value as the three interactions do not truly converge. In addition, the assumed electroweak mixing parameter, $\sin^2\theta_W = 0.375$, is at total variance with the experimental value of 0.231. Even though a 'renormalization' procedure has been devised to reduce the GU value to about 0.21 at the Z boson mass ($M_Z$), we find that, applying the renormalized value to the equations for the coupling constants leads to $\sin^2\theta_W = 0.6$ at GU, in complete contradiction to the 0.375 assumed. On a more fundamental level, although the weak and strong coupling constants are assumed to be exactly unified at GU, the incorporation of the $SU(2)_L \times U(1)$ electroweak model appears to require a modified value of the electromagnetic coupling, and so GU is not exact between the interactions but occurs only through an assumed group structure, which has no obvious explanation.

The problem with the minimal $SU(5)$ model seems to be the use of explicit fractional charges for the quarks, as would be suggested by a naïve interpretation of the phenomenology. This is what produces the unsatisfactory value for $\sin^2\theta_W$, the weak mixing angle in the GSW $SU(2)_L \times U(1)$ theory, which is calculated from

$$\sin^2\theta_W = \frac{\Sigma\, t_3^2}{\Sigma\, Q^2} . \qquad (3)$$

However, if we accept the mechanism proposed here, we will find that, although all charge-related *phenomenology* will remain exactly as assumed at present, the *gauge relations between interactions*, being of a more fundamental nature, will reflect the newly-assumed underlying structure producing the observed fractional charges. In the first place, we will obtain a new value for $\sin^2\theta_W$, for though we still have,

$$\Sigma\, t_3^2 = \frac{1}{4} \times 8 = 2 ,$$

for the weak component, with only left-handed contributions to weak isospin, from 3 colours of $u$, 3 colours of $d$, and the leptons $e$ and $\nu$, we now have

$$\Sigma\, Q^2 = 2 \times (1 + 1 + 0 + 0 + 0 + 1 + 1 + 0) = 8 ,$$

for integral charges, with both left- and right-handed contributions, rather than

$$\Sigma\, Q^2 = 2 \times \left(\frac{4}{9} \times 3 + \frac{1}{9} \times 3 + 1 + 0\right) = \frac{16}{3} ,$$

and therefore obtain

$$\sin^2\theta_W = 0.25 ,$$

rather than

$$\sin^2\theta_W = 0.375 .$$

Advanced calculations even suggest why the measured value for $\sin^2\theta_W$ might be slightly less than the theoretical one, as the presence of massive gauge bosons depresses the effective values of $1/\alpha_2$ ($\alpha_2$ being the weak fine structure constant) and $\sin^2\theta_W$ in the energy range $M_W$ - $M_z$, where they are normally measured. Theoretical plots for $\sin^2\theta_W$ against $\mu^2$ (energy$^2$) show a distinct dip at $M_W$ - $M_z$, against an overall upward trend.[26]

In standard theory, to obtain anything like the 'correct' value for $\sin^2\theta_W$, we take the first order renormalization equations for weak and strong couplings:

$$\frac{1}{\alpha_2(\mu)} = \frac{1}{\alpha_G} - \frac{5}{6\pi} \ln \frac{M_X^2}{\mu^2} \qquad (4)$$

and

$$\frac{1}{\alpha_3(\mu)} = \frac{1}{\alpha_G} - \frac{7}{4\pi} \ln \frac{M_X^2}{\mu^2} , \qquad (5)$$

and combine these with a *gauge-related* modification of the equation for the electromagnetic coupling ($1/\alpha$), based on an assumed grand unified gauge group structure:

$$\frac{1}{\alpha_1(\mu)} = \frac{1}{\alpha_G} + \frac{1}{\pi} \ln \frac{M_X^2}{\mu^2} , \qquad (6)$$

where

$$\frac{5}{3\alpha_1(\mu)} + \frac{1}{\alpha_2} = \frac{1}{\alpha} . \qquad (7)$$

From (4), (5) and (6), we obtain $M_X$ (the Grand Unified Mass scale) of order $10^{15}$ GeV, and then apply (4) and

$$\sin^2\theta_W = \frac{\alpha(\mu)}{\alpha_2(\mu)} , \qquad (8)$$

to give 'renormalized' values of $\sin^2\theta_W$ of order 0.19 to 0.21. Significantly, in this procedure, the coupling constants for the strong and weak interactions are assumed to achieve exact equalization with each other, but not with that for the electromagnetic interaction. In addition, the 'renormalization' of $\sin^2\theta_W$ is an *ad hoc* procedure, designed to give a better fit to experimental data, but without a fundamental theoretical justification. It is also *intrinsically* inconsistent, for, applying the value of $M_X = 10^{15}$ GeV to (4) - (8) to recalculate $\sin^2\theta_W$ does not produce 0.375, as initially assumed in setting up the equations, but 0.6, leading to an error of 60 %! The equations not only fail to provide answers consistent with experiment, but are also massively inconsistent with each other.

It is important to recognise that, while (4), (5) and (8) are well-established results, (6) and (7) are speculative assumptions of minimal *SU*(5), and are not supported by the experimental evidence. However, with an *independent* value for $\sin^2\theta_W$ of the right order, we can perform a much simpler calculation for $M_X$, which makes no assumptions about the group structure. In this interpretation, 0.25 becomes specifically the value for a *broken* symmetry, produced by asymmetric values of charge, whether or not it is contained within a larger group structure such as *SU*(5), and would be the value expected at the mass scale appropriate to the electroweak coupling, that is at $\mu = M_W - M_z$, and *not* the value at Grand Unification. We thus combine (4), (5) and (8) to give:

$$\sin^2 \theta_W (\mu) = \alpha(\mu) \left( \frac{1}{\alpha_3(\mu)} + \frac{11}{6\pi} \ln \frac{M_X}{\mu} \right).$$

Using typical values for $\mu = M_Z = 91.1867(21)$ GeV, $\alpha(M_Z^2) = 1/128$ (or 1/129), $\alpha_3(M_Z^2) = 0.118$ (or 0.12), and $\sin^2 \theta_W = 0.25$, we obtain $2.8 \times 10^{19}$ GeV for the Grand Unified mass scale ($M_X$). Immediately, we observe that this is of the order of the Planck mass ($1.22 \times 10^{19}$), and the result becomes even more significant, when we observe that purely first order calculations will tend to overestimate the predicted value of $M_X$. (Kounnas's two-loop calculation for the fractionally charged model reduces $M_X$ by a factor of about 0.64.[26]) Assuming, on this basis, that $M_X$ really *is* the Planck mass, we obtain $\alpha_G$ (the Grand Unified value for all interactions) $= 1/52.4$, and $\alpha_2(M_Z^2) = 1/31.5$, which is, of course, the kind of value we would expect for the weak coupling with $\sin^2 \theta_W = 0.25$.

So far, we have used only the equations for the weak and strong interactions, with $\sin^2 \theta_W = 0.25$. These equations require no modification as a result of changes in the underlying quark model, but the *U*(1) electromagnetic coupling requires alteration in the hypercharge numbers. In particular, $\binom{u}{d}_L$ changes from $1/6$ to $1/2$, while $(u^c)_L$ goes from $-2/3$ to $-1$, $-1$ or $0$, depending on the colour, and $(d^c)_L$ from $1/3$ to $0$, $0$ or $1$. The fermionic contribution to vacuum polarization is, conventionally,[27]

$$\frac{4}{3} \times \frac{1}{2} \times \left( \frac{1}{36} \times 3 + \frac{1}{36} \times 3 + \frac{1}{9} \times 3 + \frac{4}{9} \times 3 + \frac{1}{4} \times 1 + \frac{1}{4} \times 1 + 1 \right) \frac{n_g}{4\pi} = \frac{5}{3\pi},$$

where $n_g = 3$ is the number of fermion generations. However, when modified for integral charges, this becomes

$$\frac{4}{3} \times \frac{1}{2} \times \left( \frac{1}{4} \times 3 + \frac{1}{4} \times 3 + 1 + 1 + 0 + 0 + 0 + 1 + \frac{1}{4} \times 1 + \frac{1}{4} \times 1 + 1 \right) \frac{n_g}{4\pi} = \frac{3}{\pi},$$

To find out how this affects the behaviour of the electromagnetic coupling, we observe that the addition of the term $(3/\pi) \ln (M_X^2 / \mu^2)$ to $\alpha_G$ leads *directly to the coupling $\alpha$ for the electromagnetic interaction*, and not to the modified coupling $\alpha_1$, normalized to fit an overall gauge group, assumed in most Grand Unification schemes, for when $M_X = 1.22 \times 10^{19}$ GeV, $\mu = M_Z = 91.1867$ GeV, and $\alpha_G = 1/52.4$,

$$\frac{1}{\alpha_G} + \frac{3}{\pi} \ln \frac{M_X^2}{\mu^2} = 128 = \frac{1}{\alpha(\mu)}.$$

In other words, the unification at $M_X$ involves *a direct numerical equalization of the strengths of the three, or even four, physical force manifestations*, without reference to the exact unification structure; while, at Grand Unification, $C^2 = 0$ and $\sin^2\theta_W = 1$, suggesting a $U(5)$ group structure involving gravitation and a 25$^{th}$ generator. $SU(5)$ is merely the first stage of the symmetry breakdown. This unification is, indeed, far more exact than one dependent on the constants of a particular group structure, and confirms the interpretation of $\sin^2\theta_W$ as the electroweak constant for a specifically broken symmetry, taking the value of 0.25 at the energy range ($M_W$ - $M_z$) where the symmetry breaking occurs. It also allows determination of the *absolute* values of $\alpha(\mu)$, $\alpha_2(\mu)$, $\alpha_3(\mu)$, at any $\mu$, along with $\alpha_G$, for with $\sin^2\theta_W$ and $M_X$ as pure numbers, we have four equations and four unknowns. Fixing one of these (presumably $\alpha_3$) as unity at a particularly significant mass (say $m_e / \alpha$) would also establish a fix point to create an absolute scale of fermion masses, which could then be related to each other using the three fine structure constants. The result has an additional significance, in that it involves easily calculable, and experimentally testable, divergences from minimal $SU(5)$ in the three fine structure constants, particularly $\alpha$. On this model, for example, $\alpha$, at 14 TeV (the maximum energy of the proposed LHC), would be 1/118 in comparison with the 1/125 predicted by minimal $SU(5)$.

## 9 A single equation for quarks / leptons

It is possible to incorporate all the information in section 5 into a single equation for quarks and leptons (and their antistates):

$$\sigma_z \cdot (i\, \hat{\mathbf{p}}_a\, (\delta_{bc} - 1) + j\, (\hat{\mathbf{p}}_b - \mathbf{1}\delta_{0m}) + k\, \hat{\mathbf{p}}_c\, (-1)^{\delta_{1g}}\, g)$$

Here, the quaternion operators $i$, $j$, $k$ are respectively strong, electric and weak charge units; $\sigma_z$ is the spin pseudovector component defined in the $z$ direction; $\hat{\mathbf{p}}_a$, $\hat{\mathbf{p}}_b$, $\hat{\mathbf{p}}_c$ are each units of quantized angular momentum, selected *randomly* and *independently* from the three orthogonal components $\hat{\mathbf{p}}_x$, $\hat{\mathbf{p}}_y$, $\hat{\mathbf{p}}_z$.

The remaining terms are logical operators representing existence conditions. $m$ is an electromagnetic or 'weak isospin' mass unit, which becomes 1 when present and 0 when absent. The 0 condition can also be taken to mean a filled electromagnetic vacuum. $g$ represents a conjugation of weak charge units, with $g = -1$ representing maximal conjugation. If conjugation fails maximally, then $g = 1$. $g$ can also be thought of as a composite term, containing a parity element ($P$) and a time-reversal element ($T$). So, there are two ways in which the conjugated $PT$ may remain at the unconjugated value (1).

$\sigma_z$ and the three sets of logical operators define four fundamental divisions in fermionic states:

(1) $\sigma_z = -\mathbf{1}$ defines left-handed states; $\sigma_z = \mathbf{1}$ defines right-handed. For a filled weak vacuum, left-handed states are predominantly fermionic, right-handed states become antifermionic 'holes' in the vacuum.

(2) $b = c$ produces leptons; $b \neq c$ produces quarks. If $b \neq c$ we are obliged to take into account the three directions of **p** at once. If $b = c$, we can define a single direction. Taking into account all three directions at once, we define baryons composed of three quarks, in which each of $a$, $b$, $c$ cycle through the directions $x$, $y$, $z$.

(3) $m = 1$ is the weak isospin up state; $m = 0$ weak isospin down.

(4) $g = -1$ produces the generation $u$, $d$, $\nu_e$, $e$; $g = 1$, with $P$ responsible, produces $c$, $s$, $\nu_\mu$, $\mu$; $g = 1$, with $T$ responsible, produces $t$, $b$, $\nu_\tau$, $\tau$.

The weak interaction can only identify (1). This occupies the $ikE$ site in the anticommuting Dirac pentad ($ikE + i\mathbf{p} + jm$), with the $i$ term being responsible for the fermion / antifermion distinction. Because it is attached to a complex operator, the sign of $k$ has two possible values even when those of $i$ and $j$ are fixed; the sign of the weak charge associated with $k$ can therefore only be determined physically by the sign of $\sigma_z$. The filled weak vacuum is an expression of the fact that the 'ground state of the universe' can be specified in terms of positive, but not negative, energy ($E$), because, physically, this term represents a continuum state.

The strong interaction identifies (2). This occupies the $i\mathbf{p}$ (or $i\sigma.\mathbf{p}$) site and it is the three-dimensional aspect of the **p** (or $\sigma.\mathbf{p}$) term which is responsible for the three-dimensionality of quark 'colour'. A separate 'colour' cannot be identified any more successfully than a separate dimension, and the quarks become part of a system, the three parts of which have $\hat{\mathbf{p}}_a$ values taking on one each of the orthogonal components $\hat{\mathbf{p}}_x$, $\hat{\mathbf{p}}_y$, $\hat{\mathbf{p}}_z$. Meson states have corresponding values of $\hat{\mathbf{p}}_a$, $\hat{\mathbf{p}}_b$ and $\hat{\mathbf{p}}_c$ in the fermion and antifermion components, although the logical operators $\delta_{0m}$ and $(-1)^{\delta_{1g}} g$ may take up different values for the fermion and the antifermion, and the respective signs of $\sigma_z$ are opposite.

The electromagnetic interaction identifies (3). This occupies the $jm$ site in the Dirac pentad. Respectively the three interactions ensure that the orientation, direction and magnitude of angular momentum are separately conserved. Gravity (mass), finally, identifies (4).

**10 Mass**

Mass is generated when an element of partial right-handedness is introduced into an intrinsically left-handed system. So, in principle, anything which alters the signs of the terms in the expression ($i\,\hat{\mathbf{p}}_a\,(\delta_{bc} - 1) + j\,(\hat{\mathbf{p}}_b - \mathbf{1}\delta_{0m}) + k\,\hat{\mathbf{p}}_c\,(-1)^{\delta_{1g}}\,g$), or reduces any of the terms to zero, is a mass generator, because it is equivalent to introducing the opposite sign of $\sigma_z$ or a partially right-handed state. There are three main sources in the equation for producing mass. These can be described as weak isospin, quark confinement, and weak charge conjugation violation.

The two states of weak isospin produced by the term $(\hat{\mathbf{p}}_b - \mathbf{1}\delta_{0m})$ are effectively equivalent to taking an undisturbed system in the form $j\sigma_z \cdot \hat{\mathbf{p}}_b$ and of taking the same system with the added 'right-handed' term $-j\sigma_z \cdot \mathbf{1}$. In the pure lepton states, when $b = c \neq z$, and hence the weak component, $k\sigma_z \cdot \hat{\mathbf{p}}_c = 0$, the equation generates residual right-handed electron / muon / tau states, specified by $-j$, with the equivalent left-handed antistates specified by $j$. The right-handed terms may be considered as the intrinsically right-handed or non-weak-interacting parts of the fermions, generated by the presence of nonzero rest mass. (The mixing illustrates the fact that the electromagnetic interaction cannot identify the presence or absence of a weakly interacting component.) The quarks follow the same procedure as leptons in generating the two states of weak isospin, but there are no separate representations of 'right-handed' quarks, as two out of any three quarks in any baryon system will always require $c \neq z$ and $k\sigma_z \cdot \hat{\mathbf{p}}_c = 0$.

Mass is again generated by quark confinement, because each baryonic system requires quarks in which one or more of $i\sigma_z \cdot \hat{\mathbf{p}}_a$, $j\sigma_z \cdot \hat{\mathbf{p}}_b$, or $k\sigma_z \cdot \hat{\mathbf{p}}_c$ is zero. This mechanism is more likely to be relevant to composite states, such as mesons and baryons, than to 'pure' ones, such as quarks and leptons. In these cases, the mass equivalent for a zero charge would appear to be that of a fundamental unit $m_f$, from which we derive the electron mass as $m_e = \alpha m_f$. Such a mechanism has already been applied to derive the mass values associated with the baryons and mesons derived from quarks. The use of a fundamental mass unit for zero charges irrespective of origin appears to derive from the fact that these 'missing' charges are a result of a perfectly random rotation of the momentum states $\hat{\mathbf{p}}_a$, $\hat{\mathbf{p}}_b$, or $\hat{\mathbf{p}}_c$, in exactly the same manner as applies in the strong interaction to produce its linear potential; $\hat{\mathbf{p}}_a$ is, of course, actually an expression of this interaction, but $\hat{\mathbf{p}}_b$ and $\hat{\mathbf{p}}_c$ follow the identical pattern of variation.

The third mechanism for mass generation arises from the fact that the sign of the intrinsically complex $k$ term is not specified with those of the $i$ and $j$ terms. Physically, however, a filled weak vacuum requires that the weak interaction recognizes only one sign for the $k$ term when the sign of $\sigma_z$ is specified. Hence, negative values of $k\sigma_z \cdot \hat{\mathbf{p}}_c$ must act, in terms of the weak interaction, as though they were positive. Reversal of a sign is equivalent to introducing opposite handedness or mass. So, the two intrinsic signs of the $k\sigma_z \cdot \hat{\mathbf{p}}_c$ term become the source of a mass splitting between a first generation, involving no sign reversal, and a second generation in which the reversal is accomplished by charge conjugation violation. However, since charge conjugation violation may be accomplished in two different ways – either by violating parity or time reversal symmetry – there are actually two further mass generations instead of one. In addition, because the weak interaction cannot distinguish between them, the three generations represented by the quarks $d$, $s$ and $b$, are mixed, like the left-handed and right-handed states of $e$, $\mu$ and $\tau$, in some proportion related to the quark masses.

## 11 Fermion and boson masses

There are various possible relationships which might generate the spectrum of fundamental fermion and boson masses. $M_Z$ might be generated from the bosonic states of the two charge accommodation tables (equivalent to 91 GeV). The Higgs mass might be generated similarly from four tables (182 GeV); this might also represent the sum of the masses of all the fermions. (A Higgs mass generated from three tables, at 121.5 GeV, seems less likely, but is just possible.) If the fermion masses are generated in this way, and the ultimate origin of mass is in the introduction of the electric charge, to overcome symmetry violation in the weak interaction, then it is conceivable that the mass scales of the three generations of fermions are related by successive applications of the scaling factor $\alpha$. In a related way, the Cabibbo mixing between the quark generations seems to be determined by the same factor as the electroweak mixing (as we might expect) (0.23 – 0.25), and the additional mixing produced with the third generation involves terms which are the square of this factor ($\approx 0.06$).[28]

A fundamental fermion mass (probably $m_e$, via $m_f = m_e / \alpha$) seems definitely derivable from the relations between $\alpha$, $\alpha_2$ and $\alpha_3$, *without any empirical input*, but the perturbation calculations are too approximate at this stage to yield the exact value. The value produced for first order calculations using a 'unit' charge ($\alpha_3 = 1$) seems to be about 0.112 GeV (slightly above the muon mass). It is quite possible that this would be the fundamental 'unit mass' ($m_f = m_e / \alpha = 0.70$ GeV) involved in charge accommodation $SU(3)_f$, if we could do the calculation more accurately. (An alternative, but less likely, possibility would relate it to the pion mass, $2m_e / \alpha = 0.140$ GeV.) The unit nature of the strong fine structure constant at the proposed 'unit mass' would be a natural result of the strong interaction being a completely unbroken symmetry connected with an unvarying principle of 3-D rotation – an expression of 'perfect' randomness.

The 'unit mass' combined with the 'zero charge' principle, gives good values the lowest mass baryons and mesons, and might yield some information about the intrinsic mass of the $s$ quark at certain energies, as opposed to the effective masses which apply to the lighter quarks. To a lesser extent, the same may apply for the $c$ quark in the context of $SU(3)_f$. The masses of $b$ and $t$ generated this way, however, would be too low.[29]

The masses of the $d$, $s$, $b$ quarks certainly run as a result of the QCD coupling and it is generally believed that they would become identical to the respective masses of $e$, $\mu$, $\tau$ at the Grand Unification energy ($M_X$, which we have fixed at the Planck mass, $1.22 \times 10^{19}$ GeV). Kounnas's QCD calculations[30] also suggest that, at some unspecified energy ($\mu$), a relationship of the form

$$\frac{m_b(\mu)}{m_\tau(\mu)} = \alpha_3(\mu)^{12/23} \; \alpha_3(m_t)^{8/161} \; \alpha_3(M_X)^{-4/7} \; (\alpha_1(\mu), \alpha_1(M_W))^{10/41}$$

should hold. (Here, we replace Kounnas's symbolic indices with their numerical values.) If $m_t = 173.8$ GeV, $\mu = 182$ GeV, $M_X = 1.22 \times 10^{19}$ GeV, we obtain $\alpha_3(\mu) = 1.0827$; $\alpha_3(m_t) = 0.1088$; $\alpha_3(M_X) = 0.01908$. Also $1/\alpha(\mu) = 126.40$, $1/\alpha_2(\mu) = 31.846$, $1/\alpha_1(\mu) = 31.517$; $1/\alpha(M_W) = 127.9$, $1/\alpha_2(M_W) = 31.846$; $1/\alpha_1(M_W) = 32.018$. So

$$\left(\frac{\alpha_1(\mu)}{\alpha_1(M_W)}\right)^{10/41} \approx \left(\frac{\alpha(\mu)}{\alpha(M_W)}\right)^{10/41} \approx 1.003 \;.$$

From these, we derive

$$\frac{m_b(\mu)}{m_\tau(\mu)} = 2.705 \;,$$

and if $m_\tau = 1.770$ GeV, then $m_b = 4.79$ GeV. Adapting this to $m_s(\mu) / m_\mu(\mu)$, with $\alpha_3(m_c)$ replacing $\alpha_3(m_t)$, we obtain $\alpha_3(m_c) = 1/3.64$ if $m_c \approx 1.2$ GeV. Hence,

$$\frac{m_s(\mu)}{m_\mu(\mu)} = 2.832 \;,$$

and, for $m_\mu = 0.10566$ GeV, $m_s \approx 0.299$ GeV. For $m_d(\mu) / m_e(\mu)$, the perturbation expansion for $\alpha_3(m_d)$ becomes impossible if $m_d \approx 6 \times 10^{-3}$ GeV, as $\alpha_3$ then increases uncontrollably. A value of $\alpha_3(m_d) \approx 10^{12}$ would be required to generate the approximate ratio 6 / 0.511, which appears to apply.

**12 Cabibbo angle for leptons**

In Halzen and Martin's *Quarks and Leptons*, we read that 'a more involved mixing in both the *u*, *c* and *d*, *s* sectors can be used but it can always be simplified (by appropriately choosing the phases of the quark states) to the one parameter form'. They also ask: 'Why is there no Cabbibo-like angle in the leptonic sector?' And answer: 'The reason is that if $\nu_e$ and $\nu_\mu$ are massless, then lepton mixing is unobservable. Any Cabibbo-like rotation still leaves us with neutrino mass eigenstates.'[31]

To have true parity between quarks and leptons (which *must* happen if the electroweak interaction is blind to the presence of the strong charge), especially as the electroweak mixing is the mechanism which actually produces mass, we should use the same CKM matrix for quarks and leptons, and should write the mixing as between *e*, $\mu$ and $\tau$, rather than between the neutrinos. This is represented as a convention in standard theory, but there may be a reason for the convention if we attribute the introduction of mass to the presence of *e* alongside *w*. The fact that we need only one isospin state to be mixed must reflect that only one has a nonzero value of *e* for any lepton or colour of quark. If one argues that the neutrinos are mixed, then one should also argue that, by comparison with the quark sector, the other leptons should also be mixed, and that, by symmetry with the quarks, we may transform away any mixing in one of the isospin states for the leptons, and, again by symmetry, this should be that of the neutrino states.

Presumably, we could not tell physically whether it is neutrinos that are 'really' mixed or the other leptons.

**13 The Higgs mechanism**

Halzen and Martin state: 'An attractive feature of the standard model is that the same Higgs doublet which generates $W$ and $Z$ masses is also sufficient to give masses to the leptons and quarks.' They then proceed to apply this to electrons, following which they state: 'The quark masses are generated in the same way. The only novel feature is that to generate a mass for the upper member of a quark doublet, we must construct a new Higgs doublet from $\phi$.' 'Due to the special properties of $SU(2)$, $\phi_c$ transforms identically to $\phi$, (but has opposite weak hypercharge to $\phi$, namely $Y = -1$). It can therefore be used to construct a gauge invariant contribution to the Lagrangian.'[32] Significantly, the hypercharge of $(u_L, d_L) = -1$ in the integral charge model, when the charge structure matches that of the leptons and $\sigma_z \cdot \hat{\mathbf{p}}_b = -1$, but becomes 1, when $\sigma_z \cdot \hat{\mathbf{p}}_b = 0$, and the electric charge component is provided purely by the filled electromagnetic vacuum.

However, this is not so for fractional charges! There the hypercharge is an invariable 1/3 and there is no negative term: fractional charges allow only one hypercharge state, though the mechanism requires two. The necessary asymmetry introduced by integral charges is lost. It requires the invention of an arbitrary and unphysical linear combination to relate the Higgs terms to the $u$ and $d$ quark masses in the fractional model whereas the relationship a natural consequence of using integral charges. And, of course, the reason why the hypercharge must be reversed is that the transition involves a reversal of the 'electromagnetic vacuum' or background condition, from empty to full. The Higgs mechanism seems to make perfect sense of this procedure, where it is just a mathematical operation that 'works' with appropriate (unexplained) adjustments in the conventional view.

In the integral-charge model, the Higgs Lagrangian for the mass of $e$ directly transfers from the usual covariant derivative Lagrangian (it is a virtually a direct copy now applied to the Higgs doublet); and then the fermion mass Lagrangian for $d$ when the charge structure matches that of the leptons is a direct copy of that for $e$, while the fermion mass Lagrangian for $d$ when the charge structure is not lepton-like is a direct copy of that with reversed hypercharge. In the conventional model, the hypercharge for quark mass is different from the hypercharge for quarks; here it is the same. The use of the Higgs mechanism with integral charges requires no extra modelling at all, and it stems from a charge 'vacuum' (the absence of charges).

## 14 The CKM mixing

Let us imagine that, ideally, the Cabibbo mixing is 1/4 for the first generation and 1/16 for the second, and suppose that we may devise an ideal CKM matrix of the form:

$$\begin{pmatrix} 1 & 0.25 & 0 \\ 0.25 & 1 & 0.0625 \\ 0 & 0.0625 & 1 \end{pmatrix}$$

Let us now suppose that this matrix acts upon a set of lepton states $e$, $\mu$, $\tau$ to produce a mixed set $e'$, $\mu'$, $\tau'$. That is, though there is no compulsion or mechanism for leptons to be mixed in the same way as quarks, the symmetry determining the masses of $e$, $\mu$, $\tau$ requires a set of mixed states $e'$, $\mu'$, $\tau'$, such that

$$\begin{pmatrix} 1 & 0.25 & 0 \\ 0.25 & 1 & 0.0625 \\ 0 & 0.0625 & 1 \end{pmatrix} \begin{pmatrix} e \\ \mu \\ \tau \end{pmatrix} = \begin{pmatrix} e' \\ \mu' \\ \tau' \end{pmatrix}$$

Applying the principle that the fermion masses are generated through the perfectly random rotation of $\hat{\mathbf{p}}_a$, $\hat{\mathbf{p}}_b$, and $\hat{\mathbf{p}}_c$, we might expect that the intrinsic masses of the fermions are related in some way to the constant $\alpha_3$, which provides the 'unit' mass under ideal conditions. Using the accepted values for the respective masses of $e$, $\mu$ and $\tau$ at $0.511 \times 10^{-3}$, $0.10566$ and $1.770$ GeV, we obtain the respective mass values of $e'$, $\mu'$ and $\tau'$ as $0.0269$, $0.2164$ and $1.777$ GeV, with the corresponding mass ratios of $\tau' / \mu' = 8.21$ and $\mu' / e' = 8.04$. These values are essentially equal to $1 / \alpha_3$ at the energy of the electroweak splitting represented in the CKM matrix (with $\alpha_3$ possibly decreasing slightly at the higher energy required in the third generation).

So, continuing the parallel between the lepton and quark sets, we imagine that the separation between the mass values for $e'$ and $\mu'$ is determined by the 'strong' factor $\alpha_3$ (at the energy of $M_W - M_Z$), with the first generation mass being $\alpha_3 \approx 1 / 8$ times that of the second, and that the same applies to the separation between the mass values for $\mu'$ and $\tau'$. Again, the connection with $\alpha_3$ occurs through the connection between the strong interaction potential and the perfectly random rotation of the angular momentum operators, rather than due to the necessary presence of any strong charge; so, perfect randomness applied to lepton angular momentum operators has the same structure as that applied to those defined for the quarks in baryons and mesons. In principle, it is the perfectly random rotation of the angular momentum states, $\hat{\mathbf{p}}_a$, $\hat{\mathbf{p}}_b$, and $\hat{\mathbf{p}}_c$, which *determines the behaviour of the strong interaction*, with its linear potential and asymptotic freedom, and the value of its fine structure constant, $\alpha_3$, and associated unit mass; and not the strong interaction which determines the rotation of the angular momentum states.

The result of the CKM calculations seems to be that the masses of $e'$, $\mu'$, $\tau'$ are determined as though in a quark mixing, though there is no actual mixing between $e$, $\mu$ and $\tau$. In addition, the CKM matrix values will be the idealised ones as there is no 'running' aspect to the lepton masses. If, then, as is highly probable, the mass of $e$ is determined uniquely in the form of $m_e / \alpha$ for 'unit charge', then the masses of $e$, $\mu$ and $\tau$ are, in principle, determined absolutely.[33]

Identical considerations should apply to the quarks $d$, $s$, $b$ and their CKM-rotated equivalents $d'$, $s'$, $b'$. At GU their masses would be the same as those of the free fermions or leptons. However, at other energies, the mass values associated with $d$, $s$, $b$ and $d'$, $s'$, $b'$ become variable, along with the fine structure constants, and, presumably, the mixing angles. The exact CKM parameters would be similar to the idealized ones but would diverge from them according to the necessity of fulfilling such conditions, from the renormalization of $\alpha_3$, as the quark masses at measurable energies being approximately 3 times the lepton masses; and of fixing the sum of fermion masses at 182 GeV.

According to the Higgs mechanism, and our own representation, the weak isospin up state of the quarks $u$, $c$, $t$ represents a filled electromagnetic vacuum. We may therefore expect the separation of the generation masses to be determined by the electromagnetic factor $\alpha$ (at some suitable energy). The electromagnetic connection is also obvious from the origin of this mass in the term $\boldsymbol{j}\,(\hat{\mathbf{p}}_b - \mathbf{1}\delta_{0m})$. So the mass of $c$ is $\alpha$ times that of $t$, and the mass of $u$ is $\alpha$ times that of $c$. (Possibly this applies to the quark generations, rather than the individual particles.) The masses due to the weak isospin 'up' states, as is evident from the general formula for fermions, does not come from the perfectly random rotation, which determines the masses of all other states. The mass of the $t$ quark represents the maximum possible energy for a state $f / \sqrt{2}$, where $f$ is the vacuum expectation value, which is, of course, the same for quarks and free fermions (and, so is of order $3M_W$).

**Notes and References**